\documentclass[twocolumn,showpacs,aps,epsfig,nofootinbib]{revtex4}
\usepackage{graphicx}
\usepackage{epstopdf}
\usepackage{latexsym}
\usepackage{amssymb}
\usepackage[center]{subfigure}

\begin{document}

%%%%%%%%%%%%%%%%%%%%%%%%%%%%%%%%%%%%%%%%%%%%%%%%%%%%%%%%%%%%%%%
 \newcommand{\bq}{\begin{equation}}
 \newcommand{\eq}{\end{equation}}
 \newcommand{\bqn}{\begin{eqnarray}}
 \newcommand{\eqn}{\end{eqnarray}}
 \newcommand{\nb}{\nonumber}
 \newcommand{\lb}{\label}
\newcommand{\PRL}{Phys. Rev. Lett.}
\newcommand{\PL}{Phys. Lett.}
\newcommand{\PR}{Phys. Rev.}
\newcommand{\CQG}{Class. Quantum Grav.}
 %%%%%%%%%%%%%%%%%%%%%%%%%%%%%%%%%%%%%%%%%%%%%%%%%%%%%%%%%%%%%%%

\title{Effects of high-order operators in  non-relativistic  Lifshitz holography}

\author{Xinwen Wang $^{a, b}$}
%\email{Xinwen_Wang@baylor.edu}

\author{Jie Yang $^{c}$}
%\email{yangjiev@lzu.edu.cn}

\author{Miao Tian $^{b,d}$}
%\email{Miao_Tian@baylor.edu}

\author{Anzhong Wang $^{a, b}$ \footnote{The corresponding author}}
\email{Anzhong_Wang@baylor.edu}

\author{Yanbin Deng $^{e}$}
%\email{Yanbin_Deng@baylor.edu}

\author{Gerald Cleaver $^{e}$}
%\email{Gerald_Cleaver@baylor.edu}

\affiliation{$^{a}$ Institute  for Advanced Physics $\&$ Mathematics,
Zhejiang University of
Technology, Hangzhou 310032,  China\\
$^{b}$ GCAP-CASPER, Physics Department, Baylor
University, Waco, TX 76798-7316, USA\\
$^{c}$ Institute of Theoretical Physics, Lanzhou University, Lanzhou 730000, China\\
$^{d}$ School of Mathematics and Physics, Lanzhou Jiaotong University,
Lanzhou 730070, China\\
$^{e}$ EUSCO-CASPER, Physics Department, Baylor
University, Waco, TX 76798-7316, USA
}

\date{\today}

\begin{abstract}

In this paper, we study  the effects of high-order operators on the non-relativistic Lifshitz holography in the framework
of the Ho\v{r}ava-Lifshitz (HL) theory of gravity, which naturally  contains  high-order operators in order for the theory
to be power-counting renormalizble, and provides an ideal place for such  studies. In particular, we  show that
the Lifshitz space-time  is still a solution of  the full theory of the HL gravity. The effects of the high-oder operators on
the space-time itself  is simply to shift  the Lifshitz dynamical exponent. However, while in the infrared  the asymptotic
behavior of a (probe) scalar field near the boundary is similar to that  studied in the literature, it  gets dramatically
modified in the UV limit,  because of  the  presence of the high-order operators  in this regime. Then, according to the
gauge/gravity duality, this in turn affects the two-point correlation functions.

\end{abstract}


\maketitle

\section{Introduction}
\renewcommand{\theequation}{1.\arabic{equation}} \setcounter{equation}{0}

Non-relativistic gauge/gravity duality  has attracted lot of attention recently, as it may provide valuable tools to study strongly coupling systems
encountered  in condensed matter physics \cite{Sachdev}, which otherwise are not tractable with our current understanding.  If such a duality indeed
exists, instead of directly studying those  strongly coupling systems, one can study the corresponding weakly coupling systems of gravity, which are much easier to
handle, and often well within our abilities.

The non-relativistic quantum field theories (NQFT) are usually assumed to possess  either the Schr\"odinger \cite{Son} or the Lifshitz \cite{KLM} symmetry. In the latter,
the   symmetry algebra consists of the  rotations  $M_{ij}$, spatial translations $P_{i}$, time translations $H$,  and dilatations $D$. These generators satisfy the standard
commutation relations for $M_{ij}, P_{k}$ and $H$ \cite{Weinberg}, while with $D$ the relations  read,
\bqn
\lb{1.1}
&&\left[D,M_{ij}\right] = 0, \;\;\;
\left[D,P_i\right] = iP_i, \;\;\;
\left[D,H\right] = izH,  ~~~~
\eqn
where $z$ denotes the Lifshitz dynamical  exponent, and determines the relative scaling between the time and spatial coordinates \cite{Lifshitz},
\bq
\lb{1.2}
x^i \rightarrow \ell x^i, \;\;\;  t \rightarrow \ell^{z} t.
\eq
This algebra is often  called the Lifshitz algebra, as it generalizes the symmetry of Lifshitz fixed points \cite{Sachdev}.

The gauge/gravity duality requires that the space-time in the gravitational side must possess the same symmetry. However, the symmetry of a space-time  is usually defined by the existence of
Killing vectors $\zeta_{\mu}$ \cite{HE73}, satisfying the Killing equations,
\bq
\lb{1.3}
\zeta_{\mu;\nu} + \zeta_{\nu;\mu} = 0,
\eq
where a semicolon ``;" denotes the covariant
derivative with respect to the spapcetime metric $g_{\mu\nu}$. It was found that this can be realized in the Lifshitz space-time \cite{KLM},
\bq
\lb{1.4}
ds^2 \equiv g_{\mu\nu}dx^{\mu}dx^{\nu} = - r^{2z}dt^2 + \frac{dr^2}{r^2} + r^2d\vec{x}^2,
\eq
where $d\vec{x}^2 \equiv \sum_{i=1}^{d}dx^idx^i$. Then,  the Killing vectors $\zeta^{\mu}\partial_{\mu}  \equiv \left(M, P, H, D\right)$ of the above space-time, given by,
\bqn
\lb{1.5}
&& M_{ij} = - i\left(x_i\partial_j - x_j\partial_i\right),\;\;\;
P_i = - i\partial_i,\nb\\
&& H = -i \partial_t,\;\;\; D = -i\left(zt \partial_t + x^i \partial_i  - r \partial_r\right),
\eqn
produce precisely the required Lifshitz algebra, where $x_i \equiv \delta_{ij}x^j$.
The corresponding NQFT lives on the boundary  $r = \infty$.

Note that  the metric is invariant under the rescaling (\ref{1.2}), provided that $r$ is scaling as $r \rightarrow \ell^{-1} r$. Clearly,  this is non-relativistic for $z \not= 1$, and
to produce such a space-time in Einstein's theory of general relativity (GR), matter fields must be present, in order to create such a preferred  direction. In \cite{KLM}, this
was realized by two p-form gauge fields with $p = 1, 2$, and was soon generalized to other cases   \cite{Mann}.

On the other hand, to construct a viable theory of quantum gravity, Ho\v{r}ava \cite{Horava} recently proposed a theory based on the anisotropic scaling (\ref{1.2}), the so-called
Ho\v{r}ava-Lifshitz (HL) theory of   quantum gravity, and  has attracted a great deal of attention, due to its several remarkable features \cite{reviews}.
 The  HL  theory  is based on the perspective that Lorentz symmetry should appear as an emergent symmetry at long
distances, but can be fundamentally  absent at short ones \cite{Pav}. In the UV regime,  the system
 exhibits a strong anisotropic scaling between space and time, given by Eq.(\ref{1.2}). To have the theory be power-counting renormalizable, the Lifshitz dynamical exponent
 $z$ must be no less than $D$  in  the $(D+1)$-dimensional spacetime \cite{Horava,Visser}.
 At long distances, high-order curvature corrections become negligible, and  the lowest order terms     take over,  whereby the Lorentz invariance is expected to be
 ``accidentally restored."

 Since in the HL gravity the anisotropic scaling (\ref{1.2}) is built in  \footnote{It should be noted that in the HL gravity, all the spatial coordinates $\left(r, x^i\right)$ are
 scaling as $x^n \rightarrow \ell x^n$, where $n = r, i, (i = 1, 2, 3, ..., d)$. This is different from that of the metric (\ref{1.4}), in which $r$ must be scaling as
 $r \rightarrow \ell^{-1} r$, in order to keep the metric invariant. Therefore, in principle the Lifshitz dynamical exponent $z$ appearing in  (\ref{1.4}) is
 different from that considered in the HL theory:  $x^n \rightarrow \ell x^n, \;\;\;  t \rightarrow \ell^{z} t$.},   it is natural to expect that the HL gravity provides a minimal holographic dual for
non-relativistic Lifshitz-type field theories. Indeed, recently it was  showed that the Lifshitz spacetime (\ref{1.4})
is a vacuum solution of the HL gravity in (2+1) dimensions, and that   the full structure of the $z=2$ anisotropic Weyl anomaly can be reproduced  in dual field theories \cite{GHMT},
while  its minimal relativistic gravity counterpart yields only one of two independent central charges in the anomaly. This speculation has been further confirmed by the existence of
other types of  the Lifshitz spacetimes, including Lifshitz solitons \cite{SLWW,LSWW}.

 In this paper, we study another important issue: the effects of high-order operators in  non-relativistic  Lifshitz holography. Since high-order operators are necessarily  appear in the HL gravity
 in order to be power-counting renormalizable,  it provides an ideal place to study such effects.  In the framework of GR, this was studied in \cite{AMSV}, and
 found that these effects only shift the values of $z$. In this paper, we shall first show that this is true also in the HL gravity. Then, we study the effects on a scalar field and the corresponding
 two-point correlation functions. We find that, while in the infrared  the asymptotic  behavior of a (probe) scalar field near the boundary is similar to that  studied in \cite{KLM}, it  gets dramatically
modified in the UV limit,  because of  the  presence of the high-order operators  in this regime. Then, according to the gauge/gravity duality, this in turn affects the two-point correlation functions.
This is expected, as in the UV the high-order operators will dominate, and  the asymptotic behavior of the scalar field will be determined by these high-order operators.

Specifically, the paper is organized as follows: In Section II, we shall give a brief introduction to the non-projectable HL gravity in (2+1)-dimensional spacetimes, and
find out the stability and ghost-free conditions in terms of the independently coupling constants of the theory. In Section III, we show that the Lifshitz space-time (\ref{1.4}) is not only a solution
of the HL gravity in the IR limit, but also a solution of the full theory. The only difference is that the Lifshitz dynamical exponent $z$ is shifted. In Section IV, we study a scalar field propagating on the Lifshitz background (\ref{1.4}). To compare our results with the ones obtained in \cite{KLM}, in this section (and also the next)  we set $z =2$. In Section V, we calculate  the two-point correlation functions, and find their main properties in the IR as well as in the UV limit. In Section V, we present our main conclusions.

\section{ Non-projectable HL theory  in (2+1) dimensions  }

\renewcommand{\theequation}{2.\arabic{equation}} \setcounter{equation}{0}

Because of the anisotropic scaling (\ref{1.2}) [see also Footnote 1],   the gauge symmetry of the theory is broken down to the  foliation-preserving
diffeomorphism, Diff($M, \; {\cal{F}}$),
\bq
\lb{2.0a}
\delta{t} =  - f(t),\; \;\; \delta{x}^{i}  =    - \zeta^{i}(t, {\bf x}),
\eq
for which the lapse function $N$, shift vector  $N^{i}$, and 3-spatial metric $g_{ij}$, first  introduced in the  Arnowitt-Deser-Misner (ADM) decompositions \cite{ADM},    transform as
\bqn
\lb{2.0b}
\delta{N} &=& \zeta^{k}\nabla_{k}N + \dot{N}f + N\dot{f},\nb\\
\delta{N}_{i} &=& N_{k}\nabla_{i}\zeta^{k} + \zeta^{k}\nabla_{k}N_{i}  + g_{ik}\dot{\zeta}^{k}
+ \dot{N}_{i}f + N_{i}\dot{f}, \nb\\
\delta{g}_{ij} &=& \nabla_{i}\zeta_{j} + \nabla_{j}\zeta_{i} + f\dot{g}_{ij},
\eqn
where $\dot{f} \equiv df/dt,\;  \nabla_{i}$ denotes the covariant
derivative with respect to   $g_{ij}$,  $N_{i} = g_{ik}N^{k}$, and $\delta{g}_{ij}
\equiv \tilde{g}_{ij}\left(t, x^k\right) - {g}_{ij}\left(t, x^k\right)$,
 etc.

 Due to the  Diff($M, \; {\cal{F}}$) diffeomorphisms (\ref{2.0a}), one more degree of freedom appears
in the gravitational sector - a spin-0 graviton.
Using the gauge freedom (\ref{2.0a}),
without loss of the generality, one can always set
\bq
\lb{gaugeB}
N^i = 0,
\eq
for which the remaining gauge freedom is
\bq
\lb{GaugeC}
t = \hat f(t'),\;\;\; x^i = \hat\zeta^i(x').
\eq
In the rest of this section, we shall leave the gauge choice open, and in particular not restrict ourselves to the  gauge (\ref{gaugeB}).

The   Riemann and Ricci tensors $R_{ijkl}$ and $R_{ij}$ of the
2D leaves $t = $ constant   are uniquely determined
by the 2D Ricci scalar $R$ via the relations \cite{SC98},
\bqn
\lb{2.0aa}
R_{ijkl} &=& \frac{1}{2}\left(g_{ik}g_{jl} - g_{il}g_{jk}\right)R,\nb\\
R_{ij} &=& \frac{1}{2}g_{ij}R, \; (i, j = 1, 2).
\eqn

 The general action of the HL theory without the projectability condition in (2+1)-dimensional spacetimes is given by \cite{SLWW}
 \bqn
  \lb{2.1}
S &=& \zeta^2\int dt d^{2}x N \sqrt{g} \Big({\cal{L}}_{K} -
{\cal{L}}_{{V}}   +{\zeta^{-2}} {\cal{L}}_{M} \Big),
 \eqn
where $g={\rm det}(g_{ij})$, $\zeta^2 = {1}/{(16\pi G)}$, and
 \bqn
 \lb{2.2}
{\cal{L}}_{K} &=& K_{ij}K^{ij} -   \lambda K^{2},\nb\\
{\cal{L}}_{V} &=&  \gamma_0 \zeta^2  + \beta   a_{i}a^{i} + \gamma_1 R \nb\\
&&
+ \frac{1}{\zeta^{2}} \Big[\gamma_{2}R^{2}   + \beta_{1} \left(a_{i}a^{i}\right)^{2} + \beta_{2} \left(a^{i}_{\;\;i}\right)^{2} \nb\\
 && + \beta_{3} a_{i}a^{i} a^{j}_{\;\;j} + \beta_{4} a^{ij}a_{ij}\nb\\
&& +  \beta_{5}  a^ia_i R +  \beta_{6}  a^i_{\;\;i} R\Big],
 \eqn
with
 $\Delta \equiv g^{ij}\nabla_{i}\nabla_{j}$,   and
\bqn
 \lb{2.3}
K_{ij} &=& \frac{1}{2N}\left(- \dot{g}_{ij} + \nabla_{i}N_{j} +
\nabla_{j}N_{i}\right),\nb\\
a_i &=& \frac{N_{,i}}{N},\;\;\; a_{ij} = \nabla_{i}a_j.
\eqn
 ${\cal{L}}_{{M}}$ is the Lagrangian of matter fields. Then, the corresponding field equations and conservation laws are given explicitly in
 \cite{SLWW}.

 \subsection{Stability and Ghost-free  Conditions}

 It is easy to show that the Minkowski space-time
 \bq
 \lb{2.8}
 \left(\bar N, \bar{N}^i, \bar g_{ij}\right)= \left(1, 0, \delta_{ij}\right),
 \eq
is a solution of the above HL gravity with  $\gamma_0 = 0$.   Then, its linear perturbations are given by
\bqn
\lb{2.9}
\delta N &=&  n,\;\;\; \delta N_i = \partial_{i}B - S_i,\nb\\
\delta g_{ij} &=&  -2\psi\delta_{ij}    + \left(\partial_i\partial_j  - \delta_{ij}\partial^2\right)E + 2F_{(i,j)},
\eqn
where $F_{(i,j)} \equiv (F_{i,j} + F_{j,i})/2$, and
\bq
\lb{2.10}
\partial^{i}S_{i} = \partial^{i}F_{i} = 0.
\eq
It is interesting to note that in the decompositions (\ref{2.9}) no tensor mode appears in $\delta{g}_{ij}$. This is closely related to the fact that in (2+1)-dimensional
spacetimes, spin-2 massless gravitons  do not exist.

Then, the infinitesimal gauge transformations (\ref{1.4}) can be written as
\bq
\lb{2.11}
f = \epsilon(t),\;\;\; \zeta^i =\partial^i \zeta + \eta^i,\;\; (\partial_i\eta^i = 0),
\eq
under which  the quantities defined in Eq.(\ref{2.9}) transfer as,
\bqn
\lb{2.12a}
\tilde{n} &=& n + \dot{\epsilon},\;\;\;
\tilde{B} = B + \dot{\zeta},\nb\\
\tilde{E} &=& E + \zeta,\;\;\;
\tilde{\psi} = \psi - \frac{1}{2}\partial^2\zeta,\nb\\
%\lb{2.12b}
\tilde{S}_i &=& S_i + \dot{\eta}_i,\;\;\;
\tilde{F}_i = F_i + \eta_i.
\eqn
Thus, from the above we can construct three scalar and one vector gauge-invariants,
\bqn
\lb{2.13}
\Psi &\equiv& \psi + \frac{1}{2}\partial^2E,\;\;\;
\Phi \equiv B - \dot{E},\nb\\
\Upsilon &\equiv& \partial^2n,\;\;\; \Phi_i \equiv  S_i - \dot{F}_i.
\eqn

 Using the above gauge freedom, without loss of the generality, we can  set
 \bq
 \lb{2.14}
 E = 0,\;\;\; F_i = 0,
 \eq
 which will uniquely fix the gauge freedom represented by $\zeta$ and $\eta_i$, while leave $\epsilon(t)$ unspecified.
 To further study the above linear perturbations, let us consider the scalar and vector perturbations,  separately.

 \subsubsection{Scalar Perturbations}

 Under the gauge (\ref{2.14}), the remaining scalars are $n,\; B$ and $\psi$, with which it can be shown that the gravitational
 sector of the action to the second-order takes the form,
 \bqn
 \lb{2.15}
 S_g^{(2)} &=& \zeta^2\int dt d^{2}x  \bigg\{2(1-2\lambda)\dot{\psi}^{2}+2(1+2\lambda)\dot{\psi}\partial^2B\nb\\
 &&  +(1-\lambda)(\partial^2B)^{2}+\beta n \partial^2n -2\gamma_1 n \partial^2\psi  \nb\\
 &&  -\frac{1}{\zeta^{2}}\big[4\gamma_2(\partial^2\psi)^{2}+(\beta_2+\beta_4)(\partial^2n)^{2} \nb\\
 && +2\beta_6(\partial^2n)(\partial^2\psi)\big]\bigg\}.
 \eqn
Its variations  with respect to $\psi,B$ and $n$ yield, respectively,
 \bq
 \lb{2.16}
 \ddot{\psi}+\frac{1}{2}\partial^2\dot{B}+\frac{\gamma_1}{2(1-2\lambda)}\partial^2n +\frac{4\gamma_2\partial^4\psi+\beta_6\partial^4n}{2\zeta^2(1-2\lambda)}=0,
 \eq
  \bq
 \lb{2.17}
 (1-2\lambda)\dot{\psi}+(1-\lambda)\partial^2B=0,
 \eq
  \bq
 \lb{2.18}
 \beta n - \gamma_1 \psi -\frac{\beta_2+\beta_4}{\zeta^2}\partial^2n-\frac{\beta_6}{\zeta^2}\partial^2\psi=0.
 \eq

From Eq.(\ref{2.17}) we can find $B$ in terms of $\psi$, and then substituting it into
(\ref{2.15}) we obtain,
   \bqn
 \lb{2.22a}
 S_g^{(2)}&=&\zeta^2\int dt d^{2}x \Bigg\{\frac{1-2\lambda}{1-\lambda}\dot{\psi}^2 +\beta n \partial^2n -2\gamma_1 n \partial^2\psi  \nb\\
 &&  -\frac{1}{\zeta^{2}}\Big[4\gamma_2(\partial^2\psi)^{2}+(\beta_2+\beta_4)(\partial^2n)^{2} \nb\\
 &&  +2\beta_6(\partial^2n)(\partial^2\psi)\Big]\Bigg\}.
 \eqn
 Then, the ghost-free condition require
 \bq
 \lb{2.23}
\frac{1-2\lambda}{1-\lambda}\geq0,
 \eq
 that is,
  \bq
 \lb{2.24}
(i)\; \lambda>1 \;\;\;{\mbox{or}}\;\;\; (ii)\;  \lambda\leq\frac{1}{2}.
 \eq

 From Eqs.(\ref{2.16})-(\ref{2.18}), on the other hand, we can get a master equation for  $\psi$, which in momentum space can be written in the form
 \bq
 \lb{2.19}
 \ddot{\psi}_{k}+\omega_k^2\psi_k=0,
 \eq
 where
  \bqn
 \lb{2.20}
 \omega_k^2&=& \frac{1-\lambda}{1-2\lambda}\Bigg(\frac{4\gamma_2k^4}{\zeta^2}+\big(\frac{\beta_6k^4}{\zeta^2}-\gamma_1k^2\big)
 \frac{\gamma_1-\frac{\beta_6k^2}{\zeta^2}}{\beta+\frac{(\beta_2+\beta_4)k^2}{\zeta^2}}\Bigg)\nb\\
 &=& \cases{-\frac{1-\lambda}{1-2\lambda}\frac{\gamma_1^2k^2}{\beta}, & $k^2/\zeta \ll 1$, \cr
 \frac{1-\lambda}{1-2\lambda}\big(4\gamma_2-\frac{\beta_6^2}{\beta_2+\beta_4}\big)\frac{k^4}{\zeta^2}, & $k^2/\zeta \gg 1$.\cr}
 \eqn
Thus, to have the mode be stable in the infrared (IR), we must require
\bq
\lb{2.21}
\beta < 0,
\eq
while its stability condition  in the ultraviolet (UV) requires
\bq
\lb{2.22}
\gamma_2 \geq \frac{\beta_6^2}{4(\beta_2+\beta_4)}.
\eq
In the intermediate range, by properly choosing other free parameters the mode can be made always stable, and such requirement does not impose any
severe  constraints. So, in the following we do not consider it any further, and simply assume that it is always satisfied.

It should be noted that the conditions (\ref{2.24}), (\ref{2.21}) and (\ref{2.22}) are valid only for the cases $\lambda \not= 1$, for which Eq.(\ref{2.21}) tells
that $\beta$ must be strictly negative, and in particular cannot be zero.

When  $\lambda = 1$,
from Eq.(\ref{2.17}) we find that
\bq
\lb{2.22ab}
\dot\psi = 0,
\eq
that is, $\psi$ does not represent a propagative mode, and we can always set it to zero by properly choosing the boundary conditions. Then,
Eqs.(\ref{2.16}) and (\ref{2.18}) reduce to,
 \bqn
 \lb{2.16a}
&& \dot{B}- \gamma_1n -\frac{\beta_6}{\zeta^2} \partial^2n=0, \\
 \lb{2.18a}
&& \frac{\beta_2+\beta_4}{\zeta^2}\partial^2n -  \beta n  =0.
\eqn
From the last equation, we can see that $n$ does not represent a propagative mode either, and can be set to zero by properly choosing the boundary conditions.
Then, Eq.(\ref{2.16a}) yields $\dot{B} = 0$, that is, $B$ is also not a propagative mode.

Therefore, in the case $\lambda =1$ there is  no gravitational propagative mode,
similar to the relativistic case \cite{SC98}. As a result, {\em all the free parameters in this case are free, as long as the stability and ghost-free conditions are concerned}.

As a corollary, we find that  the HL theory with $\beta = 0$ is viable only when $\lambda = 1$. Otherwise, the corresponding scalar mode will become unstable, as one can see
clearly from Eq.(\ref{2.20}).

 \subsubsection{Vector Perturbations}

 Under the gauge (\ref{2.14}), the remaining vector is $S_i$, with which it can be shown that the gravitational
 sector of the action to the second-order takes the form,
\bqn
 \lb{2.25}
 S_g^{(2)} &=& -\frac{\zeta^2}{2}\int dt d^{2}x N \sqrt{g} \big(S^i\partial^2S_i\big),
 \eqn
from which we find that,
\bqn
 \lb{2.26}
 \partial^2S^i=0.
 \eqn
That is, there is no propagative vector mode in the HL gravity, even the Lorentz symmetry is violated.

In summary, the above analysis shows: {\em (i) In the case $\lambda \not=1$, only spin-0 gravitons exist in the (2+1)-dimensional non-projectable HL gravity}.
Their stability and ghost-free conditions require the independent coupling constants must satisfy the conditions of Eqs.(\ref{2.24}), (\ref{2.21}) and (\ref{2.22}).
{\em (ii) In the case $\lambda =1$, the gravitational sector of the HL gravity has no free propagation mode, similar to its relativistic counterpart}. Then,
all the free parameters in this case are free, as long as the stability and ghost-free conditions are concerned.

 \subsection{Detailed Balance Condition}

To reduce the number of the coupling constants, Ho\v{r}ava imposed the detailed balance condition \cite{Horava}. The main idea is to introduce
a superpotential $W$ on the leaves $t = $ Constant,
\bq
\lb{superpotential}
W = \int{d^2x \sqrt{g} {\cal{L}}_{W}\left(R_{ij}, a_k, \nabla_{l}\right)},
\eq
so that the potential part of the action    is given by
\bqn
\lb{LV}
\hat{\cal{L}}_{{V}}^{(DB)}  = E_{ij} G^{ijkl}E_{kl},\;\;\; E_{ij} \equiv \frac{1}{\sqrt{g}}\frac{\delta W}{\delta g^{ij}},
\eqn
where $G^{ijkl}$ denotes the generalized de Witt metric on the space of metrics, and is given by
\bqn
\lb{metricDW}
G^{ijkl} \equiv \frac{1}{2}\left(g^{ik}g^{jl} + g^{il}g^{jk}\right) - \lambda g^{ij}g^{kl}.
\eqn
Power-counting renormalizibility requires that the dimension of ${\cal{L}}_{W} $ must be greater or equal to $2d$, that is,  $[{\cal{L}}_{W}]  \ge 2d$.
Taking the lowest dimension, one can see that in (2+1)-dimensional space-times, ${\cal{L}}_{W}$ in general can be cast in the form,
\bq
\lb{LW}
{\cal{L}}_{W}  = w\left(R + \mu a_ia^i - 2 \Lambda_{W}\right),
\eq
where $w, \mu$ and $\Lambda_W$ are three coupling constants. Plugging the above into Eq.(\ref{LV}) and taking Eq.(\ref{2.0a}) into account, we find that
\bqn
\lb{ELV}
E_{ij} &=& w\left[\mu\left(a_ia_j - \frac{1}{2}g_{ij} a_ka^k\right) + \Lambda_W g_{ij}\right],\nb\\
\hat{\cal{L}}_{{V}}^{(DB)} &=& \frac{w^2}{2}\left[\mu^2\left(a_ia^i\right)^2 + 4 \left(1-2\lambda\right)\Lambda_W^2\right].
\eqn
To have a healthy IR limit, the detailed balance condition is frequently allowed to be broken softly \cite{Horava,ZWWS,BLW} by adding all the low dimensional relevant terms,
$R,\; a_ia^i,\; \Lambda$,  into $\hat{\cal{L}}_{{V}}^{(DB)}$, so that the potential is finally given by
\bq
\lb{DBSLV}
{\cal{L}}_{{V}}^{(DB)}  =2\Lambda  + \beta   a_{i}a^{i} + \gamma_1 R  + \frac{\beta_1}{\zeta^{2}}   \left(a_{i}a^{i}\right)^{2},
\eq
where $\beta_1 \equiv  w^2\mu^2/2$ and $\Lambda \equiv \gamma_0\zeta^2/2$. Comparing it with ${\cal{L}}_{{V}}$ given by Eq.(\ref{2.2}),
one can see that this is equivalent to set $\gamma_2 = 0 = \beta_n\; (2\le n \le 6)$.

\section{Lifshitz Spacetimes in  (2+1)-dimensions}

\renewcommand{\theequation}{3.\arabic{equation}} \setcounter{equation}{0}

In this section we are going to study static vacuum spacetimes  with the ADM variables given by
\bqn
\lb{3.2}
N &=& r^z f(r),\;\;\; N^i = 0,\nb\\
 g_{ij} &=& {\mbox{diag.}}\left(\frac{g^2(r)}{r^2},    r^2\right),
 \eqn
in the  coordinates $(t, r, x)$, where $z$ is the dynamical Lifshitz exponent.   Then,   we find that
\bqn
\lb{3.3}
R_{ij} &=& \frac{r g' - g}{r^2 g}\delta_i^r\delta_j^r+ \frac{r^2 \left(r g'-g\right)}{g^3}\delta_i^\theta \delta_j^\theta,\nb\\
a_{i} &=&  \frac{\left( z f +r f' \right)}{r f}\delta_i^r, \;\;\; K_{ij} = 0.
\eqn
 Inserting the above into the general action (\ref{2.1}),  for the vacuum case ${\cal{L}}_{M} = 0$, we obtain
\bq
\lb{3.4}
S_{g} = - V_x \zeta^2\int{dt dr r^{z} f g {\cal{L}}_V\Big(f^{(n)}, g^{(m)}, r\Big)},
\eq
where $V_x \equiv \int{dx},\;    I^{(n)} \equiv {d^n I(r)}/{dr^n}$, and  ${\cal{L}}_V$  is given by Eq.(\ref{A.2}). Then, it can be shown that in the present case
 there are only two independent equations, which can be cast in the forms,
\bqn
\lb{VarF}
\sum_{n=0}^3 { \left( -1\right) }^{n}\; {\frac {d^{n}}{ d{r}^{n}}}\left({\frac {\delta{\cal{L}}_{g}}{\delta{f^{(n)}}}}\right)=0, \\
\lb{VarG}
\sum_{n=0}^3 {  \left( -1 \right) }^{n}\; {\frac {
d^{n}}{d{r}^{n}}}\left({\frac {\delta{\cal{L}}_{g}}{\delta{g^{(n)}}}}\right)=0,
\eqn
where ${\cal{L}}_{g} \equiv r^{z}fg {\cal{L}}_{V}$.  In terms of $f,\; g$ and their derivatives, these two equations are given by
Eqs.(\ref{A.3}) and (\ref{A.4}).

The Lifshitz spacetime corresponds to
\bq
\lb{3.6}
f = f_0,\;\;\; g = g_0,
\eq
where $f_0$ and $g_0$ are two constant. Then, the corresponding metric can be cast in the form,
\bq
 \lb{LifshitzST}
ds^2={L^2}\left\{-\left(\frac{r}{\ell}\right)^{2z}dt^2+
\left(\frac{\ell}{r}\right)^{2} dr^2+
\left(\frac{r}{\ell}\right)^{2} d{x}^2\right\},
\eq
where   $L \equiv (f_0g_0^z)^{1/(z+1)},\; \ell \equiv (g_0/f_0)^{1/(1+z)}$. Inserting Eq.(\ref{3.6}) into Eqs.(\ref{VarF}) and (\ref{VarG}), we obtain
\bqn
\lb{3.1a}
&& 2 \zeta^2\Lambda g_0^4-\zeta^2g_0^2\left[z (2+z) \beta +2\gamma_1\right]-z^3(4+3 z)\beta_1\nb\\
&& ~~~~~~~~~~ +4\gamma_2  +z \Big[z (3+2 z) \beta _2+z \left(z^2-2\right) \beta _3\nb\\
&& ~~~~~~~~~~  -(2+z) \left(\beta _4-2\beta_5 + 2\beta_6\right)\Big] =  0,\\
\lb{3.1b}
&& 2\zeta^2\Lambda g_0^4-z\zeta^2g_0^2\left(z\beta +2\gamma_1\right)-4\gamma_2+2z\left(4\gamma_2+\beta_6\right)\nb\\
&& ~~~~~~~~~~  -z^2\Bigg\{\beta_2+3\beta_4-4\beta_5 + 4\beta_6 +z\Big[3 z \beta_1-2 \beta_2\nb\\
&& ~~~~~~~~~~  -(z-2)\beta_3 +2\beta_5\Big]\Bigg\} = 0.
\eqn

In the IR  limit, all the fourth-order terms become negligible, and  the above  equations reduce to
\bqn
\lb{3.1aa}
&& 2 \Lambda g_0^2 - \left[z (2+z) \beta +2\gamma_1\right] =  0,\\
\lb{3.1bb}
&& 2 \Lambda g_0^2 -z \left(z\beta +2\gamma_1\right) = 0,
\eqn
which have the solutions,
\bqn
\lb{LZ}
z =  \frac{\gamma_1}{\gamma_1-\beta},\;\;\;
{\Lambda} = \frac{\gamma_1^2(2\gamma_1 -\beta)}{2g_0^2(\gamma_1 - \beta)^2}.
\eqn
These are exactly what were obtained in \cite{GHMT}.

When the higher-order operators are not negligible, the sum of  Eqs.(\ref{3.1a}) and (\ref{3.1b})  yields,
\bqn
\lb{3.15}
&& \Lambda = \frac{\zeta^2\big[z\beta  + \left(1-z\right)\gamma_1\big]}{\Delta} \Bigg\{z^4\big[z\beta  - \left(1+3z\right)\gamma_1\big] \beta_1 \nb\\
&& +  z^2\left[z\beta  + \left(2z^2+z+1\right)\gamma_1\right] \beta_2  \nb\\
&& +  z^4\left[\beta  + \left(z-1\right)\gamma_1\right] \beta_3  \nb\\
&& +  z^2\left[z\left(z+2\right)\beta  + \left(1-z\right)\gamma_1\right] \beta_4  \nb\\
&& + z^3\left[\left(z+2\right)\left( z - 1\right)\beta  + 4 \gamma_1 \right] \beta_5\nb\\
&& + z\left[z\left(z+2\right)\left(z+1\right)\beta  - 2 \gamma_1(z^2+1)\right] \beta_6\nb\\
&& -  4\left[z\left(z^2 + z - 1\right)\beta  + \left(z-1\right)\gamma_1\right]\gamma_2  \Bigg\},
\eqn
where
\bqn
\lb{3.16}
\Delta &=& 2\bigg\{2 z^3\beta_1 - 2 z^2 \beta_2 -z\left(z- 3\right)\beta_6 \nb\\
&& + \left(1-z\right) \left[ z^2\beta_3 + z\beta_4 - 4\gamma_2 \right] \nb\\
&& -z\left[ 2 + z\left(z-1\right) \right]\beta_5\bigg\}^2.
\eqn

The difference of Eqs.(\ref{3.1a}) and (\ref{3.1b}), on the other hand, yields,
\bq
\lb{3.13}
 az^3+ bz^2+cz+d = 0,
\eq
where
\bqn
\lb{3.14}
a &=& -2\beta_1+\beta_3+\beta_5, \nb\\
b &=& 2\beta_2-\beta_3+\beta_4-\beta_5 + \beta_6, \nb\\
c &=& -\alpha^2(\beta -\gamma_1)-4\gamma_2 -\beta_4 + 2\beta_5 - 3\beta_6,\nb\\
d &=& 4\gamma_2-\alpha^2\gamma_1,\;\;\; \alpha \equiv \zeta g_0,
\eqn
which can be used to determine the dynamical exponent $z$ in terms of the coupling constants. In general, it has three
different solutions for any given set of the coupling constants. On the other hand, Eq.(\ref{3.13}) can be also used to determine the integration constant
$g_0$ for any given $z$ and a set of  the coupling constants. In this case, we have
\bq
\lb{3.17}
g_0^2 = \frac{az^3 + bz^2 + \hat{c}z + 4\gamma_2}{\zeta^2[\gamma_1 - (\gamma_1-\beta)z]},
\eq
where $\hat{c} \equiv -4\gamma_2 -\beta_4 + 2\beta_5 - 3\beta_6$. Clearly, for the metric to have a proper signature, $z$ has to be chosen so that
$g_0^2 > 0$ for any given set of the coupling constants $(\beta_i, \gamma_j)$.

When the fourth-order corrections are small, we can expand $z$ near its   IR fixed point, $z_0$,  given by Eq.(\ref{LZ}). Writing the fourth-order coupling constants
in the form $s = s_0 + \epsilon \hat{s}$, where  $\epsilon \ll 1$, we find that
\bqn
\lb{3.17a}
 z &=& z_0+\epsilon \delta z,\nb\\
 a &=& \epsilon(-2\hat{\beta}_1+\hat{\beta}_3+\hat{\beta}_5), \nb\\
b &=&\epsilon( 2\hat{\beta}_2-\hat{\beta}_3+\hat{\beta}_4-\hat{\beta}_5 + \hat{\beta}_6), \nb\\
c &=& c_0+\epsilon(-4\hat{\gamma}_2 -\hat{\beta}_4 + 2\hat{\beta}_5 - 3\hat{\beta}_6),\nb\\
d &=& d_0+4\epsilon \hat{\gamma}_2,
\eqn
where
$$
z_0=\frac{\gamma_1}{\gamma_1-\beta},\;\;c_0=-\alpha^2(\beta-\gamma_1),\;\;d_0=-\alpha^2\gamma_1.
$$
Thus,  to the first-order of $\epsilon$  Eq.(\ref{3.13}) yields,
\bqn
\lb{3.18ab}
&&(-2\hat{\beta}_1+\hat{\beta}_3+\hat{\beta}_5)z_0^3+( 2\hat{\beta}_2-\hat{\beta}_3+\hat{\beta}_4-\hat{\beta}_5+ \hat{\beta}_6)z_0^2\nb\\
&&+(-4\hat{\gamma}_2 -\hat{\beta}_4 + 2\hat{\beta}_5 - 3\hat{\beta}_6)z_0+4 \hat{\gamma}_2 +c_0\delta z =0,\nb\\
\eqn
from which we fin that, %It is not hard to get $\delta z$ from above equation.
\bqn
\lb{3.18aa}
\delta z&=&\frac{1}{\alpha^2(\beta-\gamma_1)^4}\Big\{\gamma _1 [\beta ^2 \left(\beta _4-2 \beta _5+3 \beta _6\right)\nb\\
&&-\beta \gamma _1 \left(-2 \beta _2+\beta _3+\beta _4-3 \beta _5+5 \beta _6\right) \nb\\
&&+2 \gamma _1^2 \left(\beta _1-\beta _2-\beta _5+\beta _6\right) ]\nb\\
&&+4 \beta \gamma _2 \left(\beta -\gamma _1\right)^2 \Big\}.
\eqn
Note that in writing the above expression, without causing any confusions, we had dropped hats from all fourth-order parameters.
To study the behavior of $z$ in the UV,  let us consider some particular cases.

\subsection{ Solutions with softly-breaking detailed balance condition}

When the softly-breaking detailed balance condition is imposed, we have $\gamma_2 = \beta_i = 0,\; (i \ge 2)$. Then, Eqs.(\ref{3.13}) and (\ref{3.15}) reduce, respectively, to
\bqn
\lb{3.18a}
&& z^3+\frac{\alpha^2}{2\beta_1}\left(\beta -\gamma_1\right)z+\frac{\alpha^2}{2\beta_1}\gamma_1=0,\\
\lb{3.18b}
&& \Lambda = \frac{ \zeta^2}{4z^2\beta_1} \left[z \beta + \left(1-z \right) \gamma_1\right]\left[z\beta -\left(1+3z\right)\gamma_1\right].\nb\\
\eqn

Eq.(\ref{3.18a}) in general has three roots, and depending on the signature of ${\cal{D}}$, the nature of these roots are different, where
\bq
\lb{3.19}
{\cal{D}} \equiv  \frac{\alpha^4}{16\beta_1^2}\left[\gamma_1^2- \frac{2\alpha^2\left(\gamma_1- \beta\right)^3}{27\beta_1}\right].
\eq
Let us consider the cases ${\cal{D}} =0,\; {\cal{D}} >0$ and $ {\cal{D}}<0$, separately.

\subsubsection{${\cal{D}} =0$}

When ${\cal{D}} = 0$, we find that
\bq
\lb{3.20}
\beta_1 =  \frac{2\alpha^2\left(\gamma_1-\beta\right)^3}{27\gamma_1^2},
\eq
and Eq.(\ref{3.18a}) has three real roots,  two of which are equal and given   by
\bq
\lb{3.21}
z_1 = \frac{3\gamma_1}{\beta -\gamma_1},\quad z_2=z_3=-\frac{3\gamma_1}{2\left(\beta -\gamma_1\right)}.
\eq
Clearly, by properly choosing $\beta$ and $\gamma_1$, they can take any real values,  $z_i \in (-\infty, \infty)$.

\subsubsection{${\cal{D}}>0$}

In this case, Eq.(\ref{3.18a}) has only one real root, which can be written as
\bq
\lb{3.22}
z = \sqrt[3]{{\cal{D}}^{1/2} -\frac{q}{2}}- \sqrt[3]{{\cal{D}}^{1/2}+\frac{q}{2}},
\eq
where $q \equiv {\alpha^2}\gamma_1/({2\beta_1})$. In this case it is clear that $z$ can also take any real values for different choices of
($\beta, \gamma_1, \beta_1$). In particular, it has an extreme at
$\beta = \gamma_1$, given by $z_m = - q^{1/3}$.

\subsubsection{${\cal{D}} <0$}

In this case, Eq.(\ref{3.18a}) has three real and different roots, given by
\bq
\lb{3.23}
z_n  =  \sqrt{\frac{2\alpha^2\left(\gamma_1-\beta \right)}{3\beta_1}} \cos\left(\theta+\frac{2n\pi}{3}\right),  (n = 0, 1, 2),
\eq
where $\theta$ is defined as
\bq
\lb{3.24}
\theta = \frac{1}{3}{\mbox{arcos}}\left[\frac{\alpha^2\gamma_1}{4\beta_1}\left(\frac{6\beta_1}{\alpha^2(\gamma_1-\beta)}\right)^{3/2}\right].
\eq
Again, similar to the last two subcases, by choosing different values of the coupling constants, we can have different values of $z_n$. For example,
taking $\alpha^2=4,\; \beta =-1, \; \beta_1=0.00001,\; \gamma_1=1$, we obtain $z_1\simeq 632.205$.

\subsection{Solutions  with ${\cal{L}}_V = {\cal{F}}(R)$}

Another interesting case is the ${\cal{F}}(R)$ models \cite{Wang}, for which we have
\bq
\lb{3.25}
{\cal{L}}_{{V}}  =  {\cal{F}}(R),
\eq
where ${\cal{F}}(R)$ can be any function of $R$ (possibly  subjected to some stability and ghost-free conditions). In particular, one can take the form,
\bq
\lb{3.25a}
{\cal{F}}(R) =  2\Lambda   + \gamma_1 R + \beta {\cal{A}}^2 + \frac{\gamma_2}{\zeta^{2}}  R^{2},
\eq
which corresponds to the potential given by Eq.(\ref{2.2}) with $\beta_i = 0, \; (i = 1, ..., 6)$, where
$ {\cal{A}}^2 \equiv a_ia^i$. Note that in writing the above expression, we had kept
the $a_ia^i$ term, in order to have a healthy IR limit for any given coupling constant $\lambda$ \cite{GHMT,SLWW}.

In this case, Eqs.(\ref{3.1a}) and (\ref{3.1b}) have the solutions,
\bqn
\lb{3.26}
z &=& 1 - \frac{\alpha^2\beta}{4\gamma_2 - \alpha^2(\gamma_1-\beta)},\nb\\
\Lambda &=& \frac{\zeta^{2}}{2\alpha^4}\left\{\alpha^2\left[z(2+z)\beta + 2\gamma_1\right] - 4\gamma_2\right\}.
\eqn

%{\large\bf (Please check this.)}

\subsection{Solutions  with ${\cal{L}}_V = {\cal{G}}({\cal{A}})$}

Similar to the last case, the function ${\cal{G}}( {\cal{A}})$ can take any form in terms of $ {\cal{A}}$.
A particular case is the potential given by Eq.(\ref{2.2}) with $\gamma_1=\gamma_2=\beta_5 = \beta_6= 0$, for which we have
\bqn
\lb{3.27}
{\cal{G}}({\cal{A}}) &=&  2 \Lambda  + \beta   a_{i}a^{i} \nb\\
&&  + \frac{1}{\zeta^{2}} \Big[\beta_{1} \left(a_{i}a^{i}\right)^{2} + \beta_{2} \left(a^{i}_{\;\;i}\right)^{2} \nb\\
 && + \beta_{3} a_{i}a^{i} a^{j}_{\;\;j} + \beta_{4} a^{ij}a_{ij}\Big].
\eqn

In this case, Eq.(\ref{3.13}) reduces to
\bq
\lb{3.28}
 az^2+ bz+c = 0,
\eq
but now with
\bqn
\lb{3.29}
a &=& -2\beta_1+\beta_3, \nb\\
b &=& 2\beta_2-\beta_3+\beta_4, \nb\\
c &=& -\alpha^2\beta  -\beta_4.
\eqn
Thus, in general there are two solutions,
\bq
\lb{3.30}
z_{\pm} = \frac{1}{2(2\beta_1-\beta_3)}\left[(2\beta_2-\beta_3+\beta_4) \pm \sqrt{D}\right],
\eq
where $D \equiv (2\beta_2-\beta_3+\beta_4)^2 + 4(\alpha^2\beta  +\beta_4)(\beta_3 -2\beta_1)$. Clearly, for $z_{\pm}$ to be real, we must assume that
$D \ge 0$.

\section{Scalar Field in The Lifshitz Spacetime}
\renewcommand{\theequation}{4.\arabic{equation}} \setcounter{equation}{0}

The action of a scalar field in the HL theory takes the form,
\bqn
\lb{3.31}
S_M &=& \int dt d^{2}x N \sqrt{g} \Bigg\{\frac{1}{2N^2}[\dot{\varphi}-N^{i}\triangledown_i\varphi]^2\nb\\
&&-V(\varphi) - {\mathcal{V}}_{\phi}^{(2)} - \frac{1}{M_*^2} {\mathcal{V}}_{\phi}^{(4)}\Bigg\},
%  - \frac{1}{2}\left[1+2V_1(\varphi)\right](\triangledown\varphi)^2\nb\\
%&&-V_2(\varphi)\left(\triangledown^2\varphi\right)^2- V_4(\varphi)\triangledown^4\varphi\Bigg\}.
\eqn
where ${\mathcal{V}}_{\phi}^{(2)}$ and ${\mathcal{V}}_{\phi}^{(4)}$ are, respectively, the second and forth order operators,  made of
$R_{ij}, \; a_i, \; \nabla_i$ and $\phi$, where
\bq
\lb{3.31aa}
\left[R_{ij}\right] = 2,\;\;\; \left[a_{i}\right] = 1 =  \left[\nabla_{i}\right],\;\;\; \left[\phi\right] = 0.
\eq
In general, they  take the forms  \cite{WWM,KK},
\bqn
\lb{3.31a}
{\mathcal{V}}_{\phi}^{(2)} &=&\underline{ \frac{1}{2}\left[1+2V_1(\varphi)\right](\triangledown_i\varphi)^2} + \epsilon_1(\phi) a_i \nabla^i\phi + \epsilon_2(\phi) a_ia^i\nb\\
&&
+ \epsilon_3(\phi) R + ...,\nb\\
{\mathcal{V}}_{\phi}^{(4)} &=& \underline{V_2(\varphi)\left(\triangledown^2\varphi\right)^2} + \underline{V_4(\varphi)\triangledown^4\varphi} + \delta_1(\phi) R_{ij}\nabla^i\phi  \nabla^j\phi
 \nb\\
&&
+  \delta_2(\phi)\left(a_i\nabla^i\phi\right)^2   +  \delta_3(\phi)R^2 + ..., %+  \delta_3(\phi)\left(a_i\nabla^i\phi\right)(\triangledown\varphi)^2 + ...,
\eqn
where $V_i$, $\epsilon_i$ and $\delta_i$ are arbitrary functions of $\phi$ only, and
 the elapsing terms are the mixed ones made of   $R_{ij},\; a_i$ and $\nabla_i\phi$. When the background is fixed, these terms always give rise to low order operators
in terms of the scalar field $\phi$. For example, the term $\epsilon_1(\phi) a_i \nabla^i\phi$  appearing in ${\mathcal{V}}_{\phi}^{(2)}$ contributes  to the equation of motion of the
scalar field  only with the first-order spatial derivative, $\nabla^{i}\left[\epsilon_1(\phi) a_i\right]$, while the term $\delta_1(\phi) R_{ij}\nabla^i\phi  \nabla^j\phi $ appearing in ${\mathcal{V}}_{\phi}^{(4)}$
contributes only with the second-order spatial derivative, $\nabla^{j}\left[\delta_1(\phi) R_{ij}\nabla^{j}\phi\right]$. In addition, the term $\delta_3(\phi)R^2$ had contributions of the
form, $\delta_3'(\phi)R^2$, which acts as a potential term once the background is fixed.
Therefore, when the space-time background is fixed, the dominant terms in the UV are only the $V_2$ and $V_4$ terms appearing in Eq.(\ref{3.31a}).
In the IR, on the other hand, their contributions must be so that the resulted action is of general covariance, in order to have a consistent theory with observations \cite{LSB}
\footnote{The only possible contributions of these terms are in the intermediate energy scales. However, the study of them in these energy scales in general are very complicated,
and are hardly carried out analytically. Thus,  in this paper we shall not consider them.}.  Therefore, in this paper, without loss of the generality,  we shall keep only the underlined $V_i(\phi)$ terms appearing in
Eq.(\ref{3.31a}) and absorb the factor $M_*^{-2}$ into $V_2(\phi)$  and $V_4(\phi)$.  Then, the Variation of the action with respect to $\varphi$ yields,
\bqn
\lb{3.32}
&&\frac{1}{\sqrt{g}}\partial_t\left[\frac{\sqrt{g}}{N}(\dot{\varphi}-N^{i}\triangledown_i\varphi)\right]=\triangledown_i\left[\frac{N^i}{N}(\dot{\varphi}-N^{k}\triangledown_k\varphi)\right]\nb\\
&& ~~~ +\triangledown^{i}[N(\triangledown_{i}\varphi)(1+2V_1)]-\triangledown^2[2NV_2(\triangledown^2\varphi)]\nb\\
&&~~~  -\triangledown^4[NV_4] -N[V^{\prime}+V_1^{\prime}(\triangledown\varphi)^2\nb\\
&&~~~ +V_2^{\prime}(\triangledown^2\varphi)^2+V_4^{\prime}(\triangledown^4\varphi)].
\eqn

To compare with the results obtained in \cite{KLM}, we first set  $L=\ell=1, \; z=2$ and  $u=1/r$. Then, the metric (\ref{LifshitzST}) becomes,
\bqn
\lb{3.33}
ds^2=-\frac{1}{u^4}dt^2+\frac{1}{u^2}(dx^2+du^2).
\eqn
In the probe limit, the backreaction of the scalar field is neglected. Hence,    taking the above space-time as the  background,
and choosing
\bqn
\lb{Vns}
V&=&m^2\varphi^2, \; V_1 = a_1,\;
V_2 = \frac{\hat{a_2}}{M_*^2}\equiv a_2,\nb\\
V_4 &=& \frac{\hat{a_4}}{M_*^2}\varphi\equiv a_4\varphi,
\eqn
where $a_n$ are constants,   we find that  Eq.(\ref{3.32}) reduces to,
\bqn
\lb{3.34}
&&u^2\partial_t^2\varphi=(1+2a_1)\left(\partial_x^2\varphi+\partial_u^2\varphi-\frac{2}{u}\partial_u\varphi\right)- \frac{2}{u^2}m^2\varphi\nb\\
&&-a_4\Big[8\partial_x^2\varphi+16\partial_u^2\varphi-\frac{32}{u}\partial_u\varphi+\frac{36\varphi}{u^2}\Big]\nb\\
&&-2u^2(a_2+a_4)(\partial_x^4\varphi+2\partial_x^2\partial_u^2\varphi+\partial_u^4\varphi).
\eqn

At the boundary $u = 0$, the scalar field takes the asymptotical form,
\bqn
\lb{3.35}
\varphi \sim u^{\vartriangle}\varphi_{1}(t,x),
\eqn
where $\vartriangle$ is one of the real roots of the equation,
\bqn
\lb{3.36}
&&(1+2a_1)(\vartriangle^2-3\vartriangle)-2m^2-a_4(16\vartriangle^2-48\vartriangle+36)\nb\\
&&-2(a_2+a_4)\vartriangle(\vartriangle-1)(\vartriangle-2)(\vartriangle-3) = 0.
\eqn
From the action (\ref{3.31}), integrating it by parts and discarding boundary terms, we find that it takes the form,
\bqn
\lb{3.42}
S_M &=& \int dt d^{2}x N \sqrt{g} \Bigg\{-\frac{\varphi}{N\sqrt{g}}\partial_t{(\frac{\sqrt{g}\dot{\varphi}}{2N})}\nb\\
&&-m^2\varphi^2+ \frac{(1+2a_1)\varphi}{2N}\triangledown_{i}(N\triangledown^{i}\varphi)\nb\\
&&-\frac{a_2\varphi}{N}\triangledown^2\left(N\triangledown^2\varphi\right)- a_4\varphi\triangledown^4\varphi\Bigg\}.
\eqn
It can be shown that both  actions (\ref{3.31}) and (\ref{3.35}) are finite for
\bqn
\lb{3.43}
\triangle\textgreater\frac{3}{2}
\eqn
with the asymptotic condition (\ref{3.35}).

In the IR,  the $V_2$ and $V_4$ terms are very small, and can be set to zero safely. In addition, in this limit the scalar field should be relativistic, so $V_1 = 0$.
Hence,  the above equation reduces to
\bqn
\lb{3.37}
&& \vartriangle^2 - 3\vartriangle - 2m^2 = 0,
\eqn
which has the solutions,
\bqn
\lb{3.38}
&& \vartriangle_{\pm} = \frac{1}{2}\left(3 \pm \sqrt{9 + 8m^2}\right).
\eqn
For
\bqn
\lb{3.39}
m^2\;\textgreater-\frac{9}{8},
\eqn
in contrast to the  case considered in \cite{KLM}, now only the solution  with $\Delta = \Delta_{+}$,
\bqn
\lb{3.40}
\varphi(u,t,x)\rightarrow u^{\triangle_+}\left(\varphi(t,x)+O(u^2)\right),
\eqn
 leads to a finite action either in the form of Eq.(\ref{3.31}) or in the one of Eq.({\ref{3.42}).

In the UV, on the other hand, the $V_2$ and $V_4$ terms dominate, and Eq.(\ref{3.36}) becomes,
\bqn
\lb{3.44}
&&(a_2+a_4)\vartriangle^4-6(a_2+a_4)\vartriangle^3+(11a_2+27a_4)\vartriangle^2\nb\\
&&-(6a_2+54a_4)\vartriangle+36a_4 = 0.
\eqn
In the case $a_4=0$, the above equation reduces to
\bqn
\lb{3.46}
\vartriangle^3-6\vartriangle^2+11\vartriangle-6=0,\; \; (a_4 = 0),
\eqn
which   has solutions
\bqn
\lb{3.47}
\triangle_1=1,\;\;\;\;\;\;\triangle_2=2,\;\;\;\;\;\;\triangle_3=3, \; (a_4 = 0).
\eqn
If we choose $a_2=-a_4$, Eq.$(4.14)$ has the double root
\bqn
\lb{3.48}
\triangle=6,\; (a_2=-a_4).
\eqn

From the above analysis, one can see that the scalar field has quite different behaviors at the boundary $u = 0$ in the two limits, IR and UV.

\section{Two-Point Correlation Functions}
\renewcommand{\theequation}{5.\arabic{equation}} \setcounter{equation}{0}

The bulk field  $\varphi(u,x)$ can be written in the  form
\bqn
\lb{5.1}
\varphi(u,t,x)=\int d^{3}x^{\prime} \varphi(0,t^{\prime},x^{\prime})G(u,t,x;0,t^{\prime},x^{\prime}).
\eqn
where $\varphi(0,t,x)$ is the scalar field on the boundary and $G(u,t,x;0,t^{\prime},x^{\prime})$  the boundary to bulk propagator.
It is easy to work in the Fourier space due to the translational invariance in t and $x$. In the Fourier space, we have
\bqn
\lb{5.2}
\tilde{\varphi}{(u,\omega,k)}={\tilde{G}{(u,\omega,k)}}{\tilde{\varphi}{(0,\omega,k)}}.
\eqn

\subsection{In the IR}

In the IR, we set $a_1=a_2=a_4=0$, Eq.(\ref{3.34}) reduces to
\bqn
\lb{5.3}
&&-u^2\partial_\tau^2\varphi=\partial_x^2\varphi+\partial_u^2\varphi-\frac{2}{u}\partial_u\varphi- \frac{2}{u^2}m^2\varphi,
\eqn
and $\tilde{G}{(u,\omega,k)}$  in Fourier space satisfies the equation,
\bqn
\lb{5.4}
\partial_u^2\tilde{G}-\frac{2}{u}\partial_u\tilde{G}-(\omega^2u^2+|k|^2)\tilde{G}=0,
\eqn
with the  boundary conditions,
\bqn
\lb{5.5}
&&(i)\; \; \tilde{G}{(0,\omega,k)}=1,\nb\\
&&(ii)\;\;  \tilde{G}{(\infty,\omega,k)}\;\; {\mbox{is finite}}. % is\;\; finite,
\eqn
Note that in writing down Eq.(\ref{5.3}), we had set $t = i \tau$. Then,
the above conditions  uniquely determine the propagator $\tilde{G}{(u,\omega,k)}$,
\bqn
\lb{5.6}
\tilde{G}(u,\omega,k)&=&\frac{2}{\sqrt{\pi}}e^{-|\omega|u^2/2}\Gamma{\left(\frac{k^2}{4|\omega|}+\frac{5}{4}\right)}\nb\\
&&\times U\left(\frac{k^2}{4|\omega|}-\frac{1}{4},-\frac{1}{2},|\omega|u^2\right),
\eqn
where $U(a,b,u)$ is the confluent hypergeometric function of the second kind.
Near $u=0, \; \tilde{G}$ is given by
\bqn
\lb{5.7}
\tilde{G}= 1-\frac{k^2}{2}u^2+\frac{8\Gamma{\left(\frac{k^2}{4|\omega|}+\frac{5}{4}\right)}|\omega|^{3/2}}{3\Gamma{\left(\frac{k^2}{4|\omega|}-\frac{1}{4}\right)}}u^3 +O\left(u^4\right).
\eqn

In the IR limit and $m=0$, the action Eq.(\ref{3.31}) yields
\bqn
\lb{5.10}
S_M^* &\equiv& \frac{i}{2}S_{M} =  \frac{1}{2} \int d\tau d^{2}x N \sqrt{g} \Big\{\frac{1}{N^2}{\varphi'}^2+ (\triangledown\varphi)^2\Big\}\nb\\
      &=&\frac{1}{2} \int d\tau d^{2}x  \sqrt{^{(3)}g} g^{\mu\nu}\partial_\mu \varphi \partial_\nu \varphi,
\eqn
where $\varphi'=\frac{\partial{\varphi}}{\partial{\tau}}$. Integrating   by parts, one can show that  the on-shell bulk action is determined by the values of the field on the boundary
\bqn
\lb{5.11}
S_M^* &=& \int d\tau dx [\sqrt{^{(3)}g}g^{uu}\varphi\partial_{u}\varphi] _\epsilon ^\infty\nb\\
      &=&  \int d\omega dk {\tilde{\varphi}(0,k,\omega)\mathcal{F}(k,\omega)\tilde{\varphi}(0,-k,-\omega)},
\eqn
where we had cut off the  space at $u=\epsilon$ to regulate the bulk action, and the ``flux factor''  $ \mathcal{F}$ is defined as
\bqn
\lb{5.12}
\mathcal{F}(k,\omega) =  [\tilde{G}(u,k,\omega)\sqrt{^{(3)}g}g^{uu}\partial_u\tilde{G}(u,-k,-\omega)]_\epsilon ^\infty. ~~~~
\eqn
Since the propagator $\tilde{G}$ vanishes at $u=\infty$, $\mathcal{F}$ only receives a contribution from the cutoff at $u=\epsilon$.
The momentum space two-point function for the operator $\mathcal{O}_\varphi$ dual to $\varphi$ is given by differentiating
Eq.(\ref{5.11}) twice with respect to $\varphi(0,k,\omega)$:
\bqn
\lb{5.13}
\langle \mathcal{O}_\varphi(k,\omega)\mathcal{O}_\varphi(-k,-\omega)\rangle=\mathcal{F}(k,\omega).
\eqn
Plugging Eq.(\ref{5.7}) into Eq.(\ref{5.12}), we pick out the leading non-polynomial piece in either k or $\omega$. This gives the correlation function, after taking  the limit  $\epsilon \rightarrow 0$,
\bqn
\lb{5.14}
\langle \mathcal{O}_\varphi(k,\omega)\mathcal{O}_\varphi(-k,-\omega)\rangle
                                                 &=&-\frac{8|\omega|^{3/2}\Gamma{(a+\frac{3}{2})}}{\Gamma{(a)}}, ~~~
\eqn
where $a\equiv \frac{k^2}{4|\omega|}-\frac{1}{4}$.
Since $\Gamma{(a\simeq 0)} \rightarrow \infty$, we find that $\langle \mathcal{O}_\varphi(k,\omega)\mathcal{O}_\varphi(-k,-\omega)\rangle \simeq 0$ as $a \rightarrow0$.
When $a \gg 1$, on the other hand, we find $\langle \mathcal{O}_\varphi(k,\omega)\mathcal{O}_\varphi(-k,-\omega)\rangle\simeq -8|\omega|^{1/2}(k^2+|\omega|)$, which
 gives rise to correlations between points only with temporal  separation.

In general,  the divergence arising as $\epsilon \rightarrow 0$ from the term proportional to $u^2$ is removed via local boundary terms
\cite{BFS,KLM}, and the terms $\mathcal{O}(u^4)$ and higher vanish as the cutoff is removed when taking the limit $\epsilon \rightarrow 0$.

\subsection{In the UV}

In the UV limit, the last term in  Eq.(\ref{3.34}) dominates, and we find that %reduces to
\bqn
\lb{5.15}
\partial_\tau^2\varphi=2a_{24}(\partial_x^4\varphi+2\partial_x^2\partial_u^2\varphi+\partial_u^4\varphi),
\eqn
where $a_{24} \equiv a_2 + a_4$. In the Fourier space, this becomes
\bqn
\lb{5.16}
\partial_u^4\tilde{G}-2k^2\partial_u^2\tilde{G}+\left(k^4+\frac{\omega^2}{2a_{24}}\right)\tilde{G}=0,
\eqn
with the same boundary condition as in Eq.(\ref{5.5}).
Then, we find that
\bqn
\lb{5.17}
\tilde{G} &=& c_1e^{-u\sqrt{\rho}(\cos\frac{\theta}{2}+i\sin\frac{\theta}{2})}\nb\\
          && +(1-c_1)e^{-u\sqrt{\rho}(\cos\frac{\theta}{2}-i\sin\frac{\theta}{2})},
\eqn
where $c_1$ is an integration constant, and 
\bq
\lb{5.17aa}
\rho \cos\theta=k^2, \;\;\;\; \rho \sin\theta=\sqrt{\frac{w^2}{2a_{24}}}.
\eq
Thus, with  $m=0$, the action (\ref{3.31}) gives rise to,
\bqn
\lb{5.20}
%S_M^{**} &\equiv&
 i S_{M} &=&  \int d\tau d^{2}x N \sqrt{g}  \Big\{\frac{1}{2N^2}{\varphi'}^2+ a_2(\triangledown^2\varphi)^2 \nb\\
 && ~~~~~~~~~~~~~~~~~~~ + a_4\phi\nabla^4\phi\Big\}\nb\\
&=&\int d\omega dk \tilde{\varphi}(0,k,\omega)\int_\epsilon^\infty du \nb\\
&& \times \Big\{\frac{\omega^2}{2}\tilde{G}(u,k,\omega)\tilde{G}(u,-k,-\omega)\nb\\
&&+a_{24}k^4\tilde{G}(u,k,\omega)\tilde{G}(u,-k,-\omega)\nb\\
&&-2a_{24}k^2\tilde{G}(u,k,\omega)\partial_u^2\tilde{G}(u,-k,-\omega)\nb\\
&&+a_2\partial_u^2\tilde{G}(u,k,\omega) \partial_u^2\tilde{G}(u,-k,-\omega)\nb\\
&&+a_4\tilde{G}(u,k,\omega)\partial_u^4\tilde{G}(u,-k,-\omega)\nb\\
&&+\frac{4a_4}{u}[\tilde{G}(u,k,\omega)\partial_u^3\tilde{G}(u,-k,-\omega)\nb\\
&&~-k^2\tilde{G}(u,k,\omega)\partial_u\tilde{G}(u,-k,-\omega)]\nb\\
&&+\frac{2a_4}{u^2}[\tilde{G}(u,k,\omega)\partial_u^2\tilde{G}(u,-k,-\omega)\nb\\
&&~-k^2\tilde{G}(u,k,\omega)\tilde{G}(u,-k,-\omega)]\Big\}\tilde{\varphi}(0,-k,-\omega)\nb\\
&=& \int d\omega dk \tilde{\varphi}(0,k,\omega) \mathcal{F}(k,\omega) \tilde{\varphi}(0,-k,-\omega),
\eqn
 where
\bqn
\lb{5.21}
\mathcal{F}(k,\omega)&=&\int_\epsilon^\infty du \Big\{\frac{\omega^2}{2}\tilde{G}(u,k,\omega)\tilde{G}(u,-k,-\omega)\nb\\
&&+a_{24}k^4\tilde{G}(u,k,\omega)\tilde{G}(u,-k,-\omega)\nb\\
&&-2a_{24}k^2\tilde{G}(u,k,\omega)\partial_u^2\tilde{G}(u,-k,-\omega)\nb\\
&&+a_2\partial_u^2\tilde{G}(u,k,\omega) \partial_u^2\tilde{G}(u,-k,-\omega)\nb\\
&&+a_4\tilde{G}(u,k,\omega)\partial_u^4\tilde{G}(u,-k,-\omega)\nb\\
&&+\frac{4a_4}{u}[\tilde{G}(u,k,\omega)\partial_u^3\tilde{G}(u,-k,-\omega)\nb\\
&&~-k^2\tilde{G}(u,k,\omega)\partial_u\tilde{G}(u,-k,-\omega)]\nb\\
&&+\frac{2a_4}{u^2}[\tilde{G}(u,k,\omega)\partial_u^2\tilde{G}(u,-k,-\omega)\nb\\
&&~-k^2\tilde{G}(u,k,\omega)\tilde{G}(u,-k,-\omega)]\Big\}.
\eqn
Plugging Eq.(\ref{5.17}) into  Eq.(\ref{5.21}), and   taking the limit $\epsilon \rightarrow 0$, we find that
\bqn
\lb{5.23}
\mathcal{F}(k,\omega)&=&4a_2c_1(1-c_1)\rho^\frac{3}{2}\sin\theta\sin\frac{\theta}{2}.
\eqn

\section{Conclusions}

In this paper, we have investigated  the effects of high-order operators on the non-relativistic Lifshitz holography in the framework of the
Ho\v{r}ava-Lifshitz (HL) theory of gravity \cite{Horava}, which contains all the required high-order spatial operators in order  to be
power-counting renormalizble. The unitarity of the theory  is also preserved, because of the absence of the high-order time operators. In this sense,
the HL gravity is an ideal place to study the effects of high-order operators on the non-relativistic gauge/gravity duality.

In particular, we have first shown that the Lifshitz space-time (\ref{LifshitzST}) is not only a solution of the HL gravity in the IR, as first shown in \cite{GHMT}
and later  rederived in \cite{SLWW}, but also a solution of the full theory. The effects of the high-oder operators on the Lifshitz dynamical exponent $z$ is simply
to shift it to different values, as these high-oder operators become more and more important, as shown  explicitly in Section III. This is similar to the case studied
in \cite{AMSV}.

In Section IV, we have studied a scalar field that has the same symmetry in the UV as the HL gravity, the foliation-preserving diffeomorphism described by
Eq.(\ref{1.4}). While in the IR the asymptotic behavior of the scalar field near the boundary is similar to that given  in the 4-dimensional spacetimes  \cite{KLM},
its asymptotic behavior in the UV gets dramatically changed, so does the corresponding two-point correlation function, as shown in Section V. This is
expected, because  the  high-order operators dominate the behavior of the scalar field in the UV. Then, according to the holographic correspondence, this in turn affects
the two-point correlation functions.

It would be important  to study the effects of high-order operators on other properties of the non-relativistic Lifshitz holography, including phase transitions
and superconductivity of the corresponding non-relativistic quantum field theories defined on the boundary. In particular, it has been suggested that inflation may be described
holographically by means of a dual field theory at the future boundary \cite{InfHolo}. This might provide deep insights to the Planckian physics in the very early universe, where
(non-perturbative) quantum  gravitational effects are expected to play an important role. Recently, a powerful analytical approximation method, the so-called {\em uniform asymptotic approximation}, 
was developed \cite{Zhu,Zhu2}, which is specially designed to study such effects in the very early universe. With the arrival of the era of the precision cosmology \cite{cosmo,GalaxS}, such effects might be
within the range of the detection of the forthcoming generation of experiments \cite{Stage4}. 

Another possible application of these high-order effects might be to Hawking radiation, where
quantum gravitational effects also become important. Previous studies of such effects   showed that the Hawking radiation is robust with respect to the UV corrections  \cite{BMPS}.  
To study them in detail, one can equally apply   the uniform asymptotic approximation method developed in \cite{Zhu}  to the studies of Hawking radiation. In particular, in the spherical background, one can  simply identify 
  the radial coordinate $r$ in the  Hawking radiation with the time variable $\eta$ used  in  the inflationary  models.
  In the inflationary models, the initial  conditions are normally the Bunch-Davies vacuum, but here in the studies of Hawking radiation they
  should be the Unruh vacuum.

 \section*{Acknowledgements}

We would like to thank Jared Greenwald for the participation in the early stage of this project. This work   was supported in part by DOE, DE-FG02-10ER41692, USA (A.W.),
Ci\^encia Sem Fronteiras, No. A045/2013 CAPES, Brazil (A.W.),
and NSFC No. 11375153, China (A.W.).

\begin{widetext}

%\onecolumngrid

\section*{Appendix A: Field Equations for Satic Spacetimes}
\renewcommand{\theequation}{A.\arabic{equation}} \setcounter{equation}{0}

From Eq.(\ref{3.3}) we find that
\bqn
\lb{A.1}
R &=& \frac{2 \left(r g'-g\right)}{ g^3} ,\nb\\
\Delta R &=& \frac{2 r}{g^7}\left[15 r^2 g'^3-r g g'\left(21 g'+10 r g''\right)
+g^2 \bigg( 6 g'+r \big(6 g''+r g^{(3)}\big) \bigg)\right] ,\nb\\
a_{ij} &=&\frac{f \left(g \left(f'+r f''\right)-r f'g'\right)-r g f'^2-z f^2 g'}{r f^2 g}  \delta_i^r \delta_j^r
+ \left(\frac{r^2 \left(z f+r f'\right)}{f g^2}\right) \delta_i^\theta \delta_j^\theta, \nb\\
a_ia^i&=& \frac{\left(z f+r f'\right)^2}{f^2 g^2}, \nb\\
a^i_{\;\; i} &=& \frac{z f^2 \left(g-r g'\right)-r^2 g f'^2+r f \left(f' \left(2 g-r g'\right)+r g f''\right)}{f^2 g^3}, \nb\\
a^{ij}a_{ij} &=& \frac{1}{f^4 g^6}\Bigg\{f^2 g^2 \left(z f+r f'\right)^2+r^2 \Big[r g f'^2+z f^2 g'  -f \Big(g\left(f'+r f''\right)-r f'g'\Big)\Big]^2\Bigg\},
\eqn
and
\begin{eqnarray}
\lb{A.2}
L_V &=&{\zeta }^{2} g_{{0}}+ \frac{1}{ g^2}\left\{ \beta  z^2 - 2 \gamma_1 + \beta  r\frac{f'}{f}\left(2z + r\frac{f'}{f} \right) \right\} + 2 \gamma_1 r \frac{g'}{g^3}\nb\\ %\\ \nonumber
&&+\frac{1}{\zeta^2 g^4 } \left\{ 2\left(2\gamma_2 -\gamma_4z^2 \right) +z^2\left(\beta_1z^2+ \beta_2 + \beta_3z + \beta_4 \right) \right. \nb\\ %\\ \nonumber
&& + \frac{1}{f}\left(  2rz\left(-2\gamma_4 + 2\beta_1z^2+ 2\beta_2 + 2\beta_3z + \beta_4 \right)f' +r^2z\left(2\beta_2 + \beta_3z\right) f'' \right) \nb\\%  \\ \nonumber
&& + \frac{1}{f^2} \left( r^2\left( -2\gamma_4 + 6\beta_1z^2+ 2\left(2-z\right)\beta_2 + z\left(5-z\right)\beta_3 + 2\beta_4 \right)\left(f'\right)^2 \right. \nb\\%\\ \nonumber
&& \left.+ 2r^3\left(2\beta_2 + \beta_3z + \beta_4\right)f''f' + r^4\left(\beta_2 + \beta_4\right)\left(f''\right)^2\right) \nb\\%\\ \nonumber
&&+ \frac{1}{f}\left(  2rz\left(-2\gamma_4 + 2\beta_1z^2+ 2\beta_2 + 2\beta_3z + \beta_4  \right)f' +r^2z\left(2\beta_2 + \beta_3z\right) f'' \right) \nb\\% \\ \nonumber
&&+ \frac{1}{f^3} \left( 2r^3\left( 2\beta_1z- 2\beta_2 - \left(z-1\right)\beta_3 - \beta_4 \right)\left(f'\right)^3 \right. \nb\\%\\ \nonumber
&&  \left. \left.- r^4\left(2\beta_2 - \beta_3 + 2\beta_4\right)f''\left(f'\right)^2 \right)+ r^4\frac{(f')^4}{f^4}\left(\beta_1 + \beta_2 -\beta_3 + \beta_4\right)\right\} \nb\\% \\ \nonumber
&&+\frac{1}{\zeta^2 g^5 } \left\{ \left( 2r\left(-4\gamma_2 +6 \beta_6  + z^2\gamma_4 \right)- rz^2\left(2\beta_2+ z\beta_3 \right) \right)g' + 2r^2 \beta_6 \left( 6 g'' + r g^{(3)}\right) \right.
\nb\\%  \\ \nonumber
&& + \frac{g'}{f}\left(  r^2z\left(4\gamma_4 - 6\beta_2-3\beta_3r -2\beta_4 \right)f' -2r^3z\left(\beta_2 + \beta_4\right) f'' \right) \nb\\% \\ \nonumber
&& + \frac{g'}{f^2} \left( r^3\left( 2\gamma_4 + 2\left(z-2\right)\beta_2- 3z\beta_3 + 2\left(z-1\right)\beta_4 \right)\left(f'\right)^2 \right. \nb\\%\\ \nonumber
&& \left.\left.- 2r^4\left(\beta_2 +  \beta_4\right)f''f'\right) + r^4\frac{\left(f'\right)^3}{f^3}\left(2\beta_2 -\beta_3+2 \beta_4\right)g'\right\} \nb\\% \\ \nonumber
&& + \frac{1}{\zeta^2 g^6} \left\{ r^2\left( 2\left(2\gamma_2- 21\beta_6 \right) + z^2 \left(\beta_2 + \beta_4 \right)\right)\left(g'\right)^2 \right. \nb\\%\\ \nonumber
&& \left.  -20\beta_6  r^3 g''g' + r^3\frac{f'}{f}\left(g'\right)^2\left( \beta_2 + \beta_4\right)\left(2z+r\frac{f'}{f}\right) \right\}  %\nb\\% \\ \nonumber &&
+30 \beta_6  r^3\frac{(g')^3}{\zeta^2 g^7}. % \\ \nonumber
\end{eqnarray}
Then, the field equations (\ref{VarF}) and  (\ref{VarG}) take the forms,
\begin{eqnarray}
\lb{A.3}
0 &=& -r^z\zeta^2 \gamma_0g + \frac{r^z }{g}\left\{ 2\gamma_1 +r \beta  \left( 2(z+2) + \left[2z + 4 - r\frac{f'}{f} \right]\frac{f'}{f} + 2rf'' \right)\right\}
 -\frac{r^z}{ g^2 }\left\{2\gamma_1 + 2 \beta  r\left( z + r \frac{f'}{f} \right)  \right\} g'  \nb\\
&&+ \frac{r^z}{\zeta^2  g^3}\left\{  -4\gamma_2- 2z(z+2)\gamma_4 + z^3(3z+4)\beta_1 - z^2(2z+3)\beta_2 - z^2(z^2-2)\beta_3+ z(z+2)\beta_4  \right.\nb\\
&&  + \frac{2r}{f}\left( \left[ -2(z+2)\gamma_4 + 6z^2(z+3 )\beta_1 - 2(2z^2 + 5z + 2)\beta_2 \right.  +2z(z^2-2z+1)\beta_3  - (z^2+3z+2) \beta_4 \right]f' \nb\\
&&  + r\left[ -2\gamma_4 + 6 \beta_1 z^2 - (z^2+ 11z+ 14)\beta_2 -2z(z-1) \beta_3 - (z^2+8z+13) \beta_4 \right]f''  \nb\\
&&  \left.- 2r^2(4+z)\left[   \beta_2 + \beta_4 \right]f^{(3)} - r^3\left[   \beta_2 + \beta_4 \right]f^{(4)} \right) \nb\\
&&  + \frac{r^2}{f^2}\left(2 \left[  \gamma_4 + 3z(z+6)\beta_1  + (z^2+ 6z+4)\beta_2 -(z^2+ 6z-3) \beta_3 + (z^2+6z+ 8) \beta_4 \right](f')^2\right.  \nb\\
&&  + 4r\left[  12\beta_1z  + 4z(z+5)\beta_2 -2(2z-1) \beta_3 + (2z+13) \beta_4\right]f''f' \nb\\
&&  \left.+4r^2\left[   \beta_2 + \beta_4 \right]f^{(3)}f' + 3r^2\left[   \beta_2 + \beta_4 \right](f'')^2 \right)
  - \frac{4r^3}{f^3}\left( \left[  (3z-4) \beta_1 + (z+2)\beta_2- (z-2)\beta_3  +(z+3) \beta_4 \right]f'\right. \nb\\
&&   \left.- r\left[   3\beta_1  -2\beta_2 - \beta_3  -2 \beta_4 \right]f''\right)(f')^2
  - \frac{3r^4}{f^4}\left( 3\beta_1 -\beta_2 -\beta_3 - \beta_4 \right) (f')^{4}  \nb\\
&&+ \frac{r^z}{\zeta^2  g^4}\left\{ \left( 2r\left[ 4\gamma_2 -6 \beta_6   - 6 \beta_1  z^3 + z(z^2+ 9z+ 6)\beta_2  -z^2(3-2z) \beta_3 + z(z^2 + 4z + 1)\beta_4 \right]g' \right.\right. \nb\\
&& \left. +2r^2\left[- 6\beta_6  +2\gamma_4z +2z(z+3) \beta_2  +  z(2z+5) \beta_4 \right]g''  + 2r^3\left[ -\beta_6  + z(\beta_2  + \beta_4) \right]g^{(3)}\right) \nb\\
&&  + \frac{1}{f}\left( \left( 2r^2\left[ 2(6+z)\gamma_4 - 18 \beta_1  z^2 + 3(z^2+9z+ 8)\beta_2 +6z(z-1) \beta_3  +(3z^2+17z+18) \beta_4 \right]f' \right. \right. \nb\\
&&   \left. +2r^3\left[ 2\gamma_4 +3(10+3z) \beta_2   +(29+9z) \beta_4 \right]f''  + 12r^4\left[ \beta_2  + \beta_4 \right]f^{(3)}\right)g'  \nb\\
&&  + 2r^3\left( \left[ 2\gamma_4 + 2(2z+5) \beta_2  + (4z+9) \beta_4 \right]g''+  2r\left[ \beta_2 + \beta_4 \right]g^{(3)}\right)f' \nb\\
&& \left.+8r^4\left[   \beta_2 + \beta_4 \right]f''g'' \right)  + \frac{1}{f^2}\left( \left( -2r^3\left[ \gamma_4 + 18 \beta_1z + 6(2+z)\beta_2 +3(1-2z) \beta_3  +2(8+3z) \beta_4 \right]f' \right. \right. \nb\\
&&  \left. -18r^4\left[ \beta_2   + \beta_4 \right]f'' \right)f' g'    \left. -4r^4 \left[\beta_2 + \beta_4 \right]\left(f'\right)^2g'' \right)
 \left. - \frac{4r^4}{f^4} \left( 3\beta_1-2\beta_2 -\beta_3 -2\beta_4 \right)\left(f'\right)^3g'  \right\} \nb\\
&&+\frac{r^z}{\zeta^2g^5}\left\{ \left(- r^2 \left[ 4\gamma_2-42\beta_6 +16\gamma_4z +z(42+15z) \beta_2 +  z(34+15z)\beta_4\right]g'  \right. \right. \nb\\
&& \left.+20 r^3\left[\beta_6  + z(\beta_2 + \beta_4) \right]g''\right)g'   +\frac{2}{f} \left( \left( -r^3\left[ 8\gamma_4 +(36+15z)\beta_2 +  (32+15z)\beta_4\right]f' -30 r^4 \left[\beta_2 + \beta_4 \right]f'' \right)\left(g'\right)^2\right.\nb\\
&& \left.  -20 r^4\left[\beta_2 + \beta_4 \right]f' g'g'' \right)    \left. + \frac{15r^4}{f^2} \left(\beta_2 + \beta_4 \right)\left( f'\right)^2\left( g'\right)^2 \right\}
+ \frac{30 r^3}{g^6}\left\{-\beta_6 +\left[\beta_2 + \beta_4 \right]\left(z + r\frac{f'}{f}\right)\right\}\left(g'\right)^3, \\
\lb{A.4}
0 &=& -r^z\zeta^2f\gamma_0 +\frac{r^z }{ g^2 }f\left(  2\left[z+\frac{f'}{f}\right]\gamma_1+\left[ z^2 + 2zr \frac{f'}{f} + r^2 \frac{\left(f'\right)^2}{f^2} \right]\beta   \right)  \nb\\
&&
+ \frac{r^z}{\zeta^2  g^4}\left\{ \left( 4(1-2z) \gamma_2 -2z(z^2-1)\beta_6  +2z^2(z-2)\gamma_4 \right.   +3 \beta_1  z^4 + z^2(2z-1)\beta_2  -z^3(z-2) \beta_3 + 3\beta_4 z^2 \right) f  \nb\\
&& + 2r\left( \left[ -4 \gamma_2 +3z(z+1)\beta_6 + z(3z-2)\gamma_4 + 6 \beta_1 z^3 \right.  -4\beta_2z^2 - z^2(2z-3)\beta_3 - z(z-1) \beta_4 \right]f' \nb\\
&&  + r\left[ +3(z+1)\beta_6  + 2\gamma_4z -z(z+3)\beta_2 +  z(z+4) \beta_4 \right]f''    \left. + r^2\left[  \beta_6  - z(\beta_2 + \beta_4) \right]f^{(3)} \right)  \nb\\
&&  + \frac{r^2}{f}\left( 2\left[ \gamma_4z + 9 \beta_1z^2 + z(z-2)\beta_2- 3z(z-1)\beta_3  +z(z+2) \beta_4 \right]\left(f'\right)^2\right.   \nb\\
&&  + 2r^3\left(\left[  2\gamma_4+(z-2) \beta_2  + (z-3)\beta_4 \right]f''- r\left[\beta_2 +\beta_4 \right] f^{(3)}\right)f'    \left.+r^2\left[   \beta_2 + \beta_4 \right]\left(f''\right)^2 \right)  \nb\\
&&  +\frac{2r^3}{f^2}\left( \left[ -\gamma_4+ 6\beta_1z - (2z-1) \beta_3  +2 \beta_4 \right]f' +r\left[   \beta_2  + \beta_4  \right]f''\right)\left(f'\right)^2  \nb\\
&&  \left. + \frac{r^4}{f^3}\left[3\beta_1 - \beta_2 -\beta_3 - \beta_4 \right]\left(f'\right)^4 \right\}
+ \frac{r^z}{\zeta^2  g^5}\left\{ 2r\left( (z+2)\left[ 4\gamma_2 -2z\beta_6  + z^2( \beta_2  + \beta_4) \right]g' \right.\right. \nb\\
&& \left. +r\left[ 4\gamma_2 -  2\beta_6 z + z^2 (\beta_2  + \beta_4) \right]g'' \right)f    +2r^2\left(\left[ 4\gamma_2- 2(3+2z) \beta_6  +3z(z+2)(\beta_2 +\beta_4) \right]f' \right.  \nb\\
&&    \left. + 2r\left[-\beta_6 +z(\beta_2 + \beta_4) \right]f'' \right)g' + 4r^3 \left[ -\beta_6  + z(\beta_2 +\beta_4)\right]g'' f'  \nb\\
&& + \frac{2r^3}{f}\left( \left( (z+4)\left[  \beta_2 +\beta_4 \right]f' +2r\left[ \beta_2 +\beta_4 \right]f'' \right)f' g'
\left. +r \left[\beta_2 + \beta_4 \right]\left(f'\right)^2g'' \right)- \frac{2r^4}{f^2} \left[ \beta_2 + \beta_4 \right]\left(f'\right)^3g'  \right\} \nb\\%%
&&+\frac{5r^2r^z}{\zeta^2g^6}\left\{  \left[ -4\gamma_2 +2 \beta_6 z - z^2(\beta_2 + \beta_4)\right] -2r\left[\beta_6  - z(\beta_2 + \beta_4) \right]f'
 -\frac{r^2}{f} \left[ \beta_2 +\beta_4\right]\left(f'\right)^2  \right\}\left(g'\right)^2.
\end{eqnarray}

\end{widetext}
 %%%%%%%%%%%%%%%%%%%%%%%%%%%%%%%%%%%%%%%%%%%%%%%%%%%%%%%%%%%%%%%%%%%

\end{document}